      [1999/12/01 v1.4c Il Nuovo Cimento]
\documentclass{cimento}

\usepackage{graphicx}  
\usepackage{amsmath, amsthm, amssymb}
\title{Measurements of the top quark mass at CDF}
\author{N. van Remortel~\from{ins:x}\from{ins:y}, on behalf of the CDF collaboration}

\instlist{\inst{ins:x} Universiteit Antwerpen, Groenenborgerlaan 171 B-2020 Antwerpen, Belgium
  \inst{ins:y} Dept. of Physics and Helsinki Institute of Physics, Gustaf H\"{a}llstr\"{o}minkatu 2, FI-00014 University of 
               Helsinki, Finland}
\PACSes{\PACSit{12.15Lk, 14.65Ha}{Conference proceedings}
}
\begin{document}

\maketitle

\begin{flushright}
\end{flushright}

\begin{abstract}
We review the most recent measurements of the top quark mass using data collected by the CDF experiment at the Tevatron 
1.96 TeV $\rm{p\bar{p}}$ collider. The mass measurements are performed in all main decay modes of the produced $\rm{t\bar{t}}$ pairs 
using integrated luminosities up to 2fb$~^{-1}$. In most channels the total uncertainty is dominated by systematic effects, most of which 
are currently being revised. The precise measurement of the top quark mass is one of the Tevatron's main and long lasting legacies. 
Besides serving as a benchmark measurement at the LHC collider, it will serve as a consistency check of the Standard Model in case a 
a neutral CP even Higgs boson is found, either at the Tevatron or at the LHC.   
\end{abstract}

\section{Introduction}
The mass of the top quark is an important free parameter of the Standard Model and is of the order of the electroweak symmetry breaking 
scale. Since virtual top quarks are involved in higher-order electroweak processes, a measurement of the top quark mass serves as a 
constraint on the mass of the SM Higgs boson~\cite{electroweak} and on particles that are predicted by theories that extend the 
SM~\cite{susy}. At the Large Hadron Collider (LHC), the top quark mass will serve as a benchmark measurement and a calibration tool for jets with high 
transverse momentum.

Within the CDF experiment, top mass analyses are generally performed separately on each of the three main final state topologies that depend on the decay of the 
W daugther bosons. Each final state is, in addition, characterized by the presence of two b-quark jets from the top decays. Most analysis use this signature in 
their event selection by requiring at least one jet to be tagged as a b-quark jet. The dilepton channel (excluding tau's) contributes to 5\% of all final states 
and is characterised by the presence of two isolated leptons and a large amount of missing transverse energy due to the escaping neutrinos. The lepton+jet channel 
contributes 30\% and has a clean signature characterized by two light quark jets with an invariant mass close to the W boson mass and an isolated lepton combined 
with missing transverse energy. The all-hadronic channel contains, in addition to 2 b-jets, four light quark jets from the decays of both W's. Its branching 
fraction is 45\%, but it also suffers from overwhelming QCD backgrounds and combinatorial ambiguities when assigning jets to top quark decays.

CDF uses a fairly standardized procedure to estimate the backgrounds in each channel~\cite{ljetstemp,allhadsel}. It relies both on Monte Carlo simulation information and data itself.
The top quark mass is generally measured in each decay channel by at least two techniques: the template approach and the matrix element technique. Each analysis 
verifies its performance, biases and statistical error estimates using the PYTHIA V6.2~\cite{pythia} generator and subsequent fragmentation model tuned to CDF data. 

In its continuous pursuit of improving the top mass measurement accuracy, the CDF experiment has come op with many new ideas, some of which will be demonstrated in 
this review. The jet energy scale (JES) uncertainty used to dominate the systematic uncertainty in nearly all analyses. Now, every analysis that contains at least 
one hadronically decaying W boson, employs an in-situ measurement of the JES, thus absorbing a large fraction of the systematic uncertainty in the statistical 
uncertainty which scales as the square root of the integrated luminosity. Other ideas consist of including the correlation between the top qaurk mass and the 
$\rm{t\bar{t}}$ production cross section in one mass analysis~\cite{tuula}, a combination of several final states in one joint mass likelihood~\cite{ljetstemp}, and the use of observables that 
are less sensitive to dominating systematic effects, such as the decay length of the b hadrons fron top decays~\cite{incandela} or the transverse momentum spectrum of the final 
state leptons~\cite{greeks}. Since the precision of most measurements is currently limited by systematic uncertainties, a large efort is ongoing to revise existing and study 
possible new sources of systematic effects in the top quark mass determination.

\section{Constraining the JES with an in-situ measurement}
\begin{figure}
\begin{center}
	\begin{tabular}{cc}
	\includegraphics[width=6.8cm, clip=]{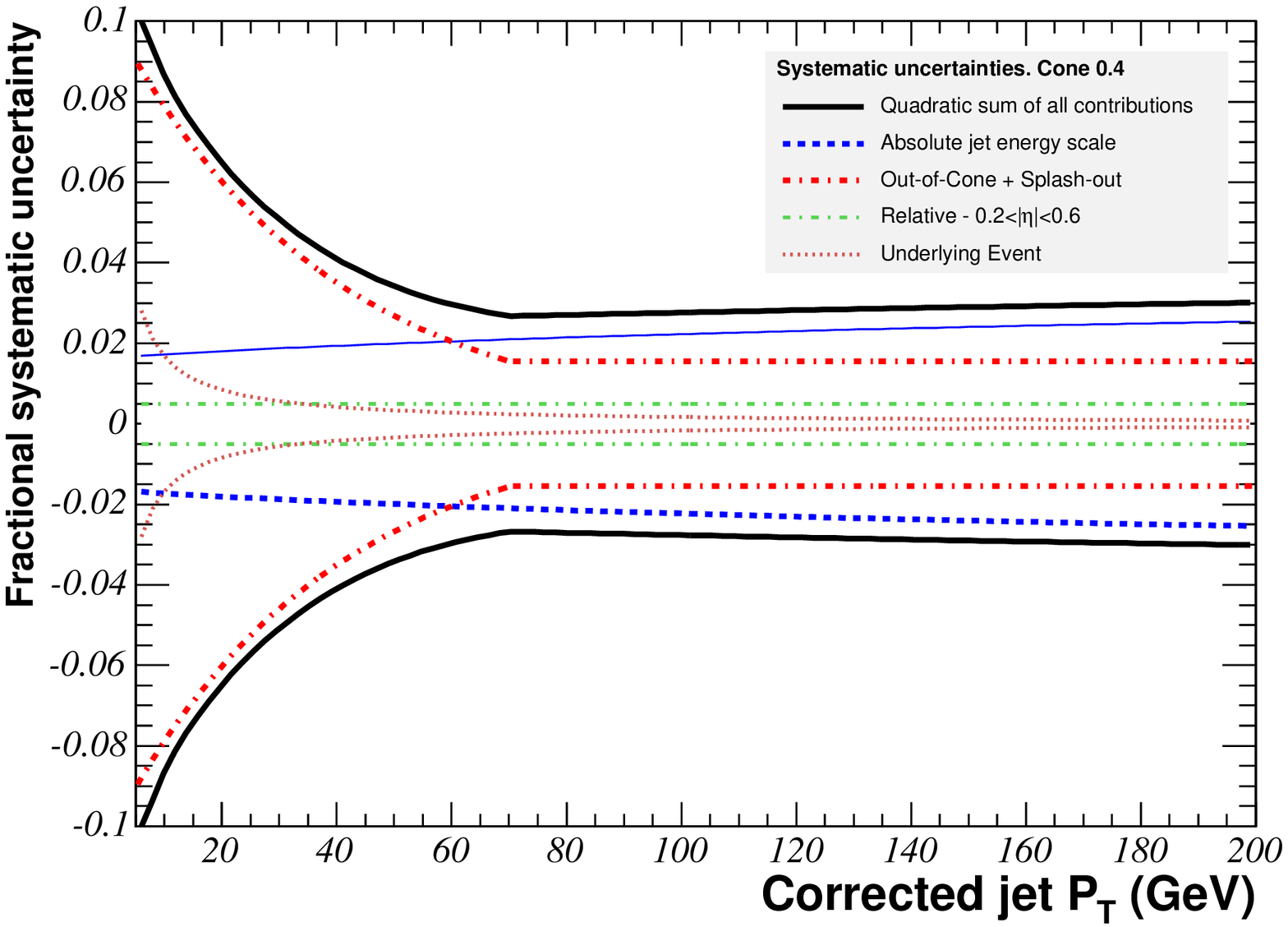}
&	\includegraphics[width=6.8cm, clip=]{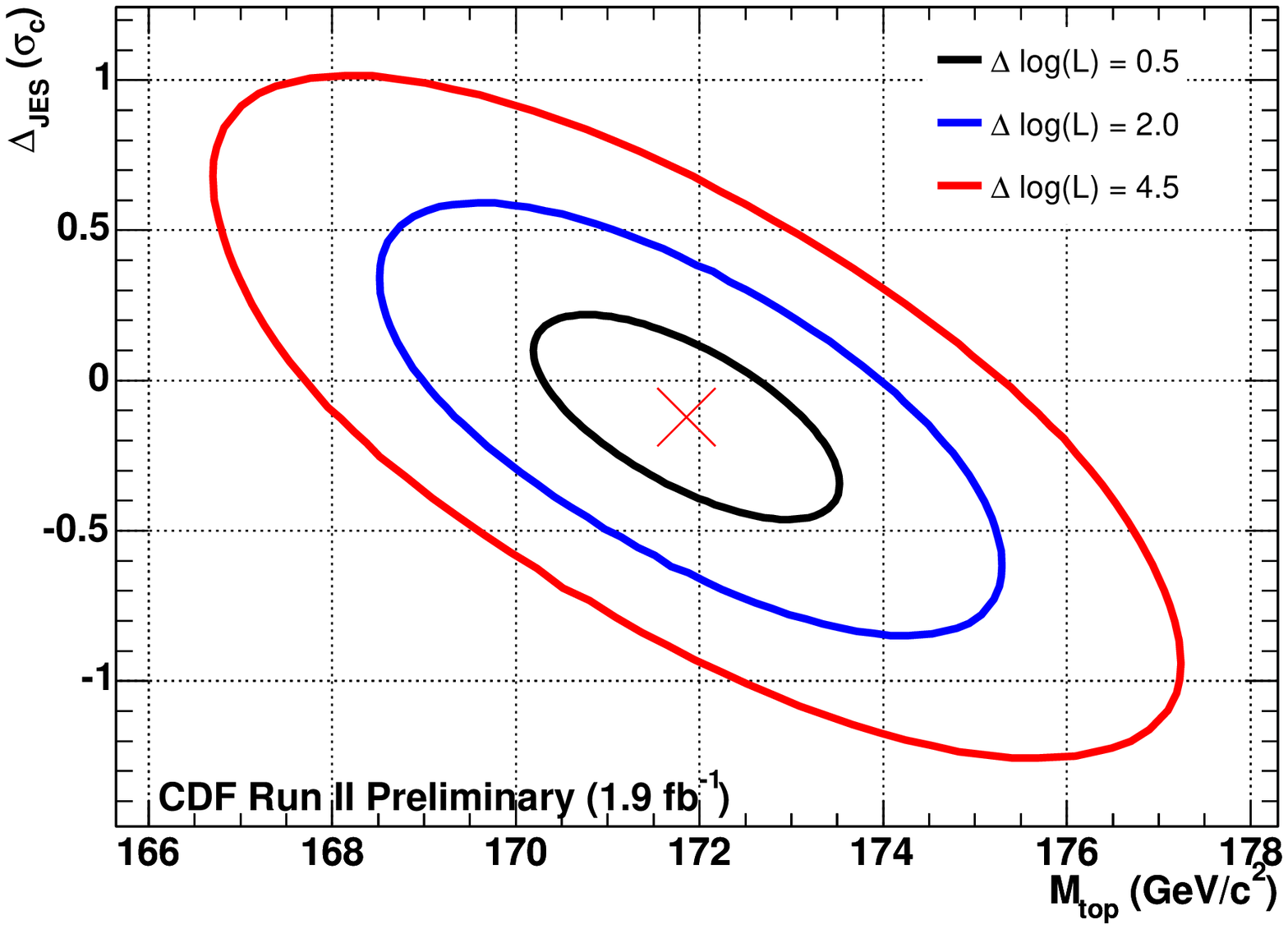}
	\end{tabular}
\caption{(left)The fractional systematic uncertainty due to the JES correction levels as function of the jet transverse momentum. The total uncertainty is taken as the sum in quadrature of all individual contributions.  (right) The log-likelihood contours for a combined fit of  $M_{top}$ and $\Delta_{JES}$ to the combined dilepton and lepton+jets dataset of 2fb$^{-1}$ of 
data collected by CDF.}
\label{JES1}
\end{center}
\end{figure}

In most CDF top mass analyses, the largest contribution to the total systematic uncertainty is due to uncertainties in the jet energy measurements. The raw jet energy scale is the result of a multi-step correction procedure~\cite{jesnim} to convert the measured transverse jet energy to the expected transverse energy of the parton corresponding to the jet. The corrections, assessed using data and simulation of the CDF detector, include corrections for the response inhomogeneity in $\eta$, contributions from multiple interactions, the non-linearity of the calorimeter response, the underlying event and the energy flow out of the jet cone. Each of those corrections has a corresponding fractional uncertainty, $\sigma_{JES}(p_T)$ which can be parameterized as function of the transverse momentum, $p_T$, of the jet, as is shown in Fig.~\ref{JES1}. 
By scaling all jet energies with a factor proportional to the average ralative uncertainty shown in Fig.~\ref{JES1} to obtain
\begin{equation}
E_{jet}=E_{meas}(1+\Delta_{JES}\sigma_{JES}(p_T)),
\end{equation}
we introduce an additional parameter, $\Delta_{JES}$,  that will be measured from data, together with the top quark mass, $M_{top}$. Besides affecting the reconstructed top quark mass in each event, this parameter mainly affects the invariant mass of the dijet system originating from a hadronically decaying W boson and will be constrained by the well know mass of the W boson. 

\section{Template methods for mass reconstruction}
\label{sect:templ}
Templates are used as a simple, robust and computationally fast method to extract the top quark mass from data. 
One or more variables that have some sensitivity to the top quark pole mass, $M_{top}$, are identified and their probability density 
is obtained by simulating a large number of $t\bar{t}$ signal events, combined in the right proportions with all possible background 
processes that are either ontained from Monte Carlo simulation, or (partially) extracted from data control regions.
These pdf's serve as templates for a series of discrete generator input values of the top quark mass, and $\Delta_{JES}$ 
in case of  an in-situ measurement of the JES. 
A maximum likelihood estimator is finally obtained by comparing the spectra in data with the templates.
In many analyses, the reconstructed top mass in each event is templated due to its large correlation with the true top quark mass.
This quantity is obtained by performing a $\chi^2$ minimization to fit the parton momenta of the $\rm{t\bar{t}}$ daughters assuming a decay
of the $\rm{t\bar{t}}$ pair into two W bosons and two b-quarks. Both sets of the W decay daughters are constrained to have the invariant 
mass of the W boson, and both Wb systems are required to have the same invariant mass. Again, for an in-situ measurement of the JES, 
the first constraint is removed and the invariant mass of the jet pair that is closest to the world avarege W boson mass is chosen as additional 
template variable. Combinatorial ambiguities are resolved by retaining only the reconstructed mass for the combination with the smallest $\chi^2$ 
value. Possible variants of this method can include techniques to smoothe and parameterize the templates, inclusion of more than one combination 
per event, or additional cuts on the $\chi^2$ minimisation outcome.

An interesting new idea that has been implemented recently is the measurement of the top quark mass by means of templates simultaneously in the dilepton 
and lepton+jets channel~\cite{ljetstemp}. As such, the measurement of the JES from the lepton+jets topology is directly fed into the measurement of the top quark mass in the dilepton channel. In addition to an intrinsic treatment of correlations in systematic uncertainties between both channels, one can directly combine likelihood curves instead of assuming Gaussian measurements in a posterior combination. This analysis measures the top quark mass to be 
$M_{top}=171.9 \pm 1.7 (stat. + JES) \pm 1.0 (syst.) GeV/c^2$ and the final two-dimensional log-likelihood constour as function of   $M_{top}$ and $\Delta_{JES}$ is shown in Fig.\ref{JES1}.

In addition to this standard approach, templates can be made from observables that are sensitive to the true top quark mass, but less sensitive to dominating 
systematic effects. Several interesting template analyses employ the shape of the transverse momentum spectrum of leptons originating from decaying W daughters 
from $\rm{t\bar{t}}$ decays in order to reduce the sensitivity to jet-related systematics~\cite{greeks}. Another method also utilizes the decay length of B hadrons in jets originating from  $\rm{t\bar{t}}$ decays as an {\it orthogonal} way to probe the top quark mass~\cite{incandela}. Finally there is also an analysis that does not rely on the explicit identification of isolated high $p_T$ leptons in the event selection, but rather on the missing transverse energy signature, which increases the sensitivity to W daughters decaying into tau leptons~\cite{jetmet}.

\section{Matrix element methods}
\begin{figure}[t]
\begin{center}
	\begin{tabular}{cc}
	\includegraphics[width=6.8cm, clip=]{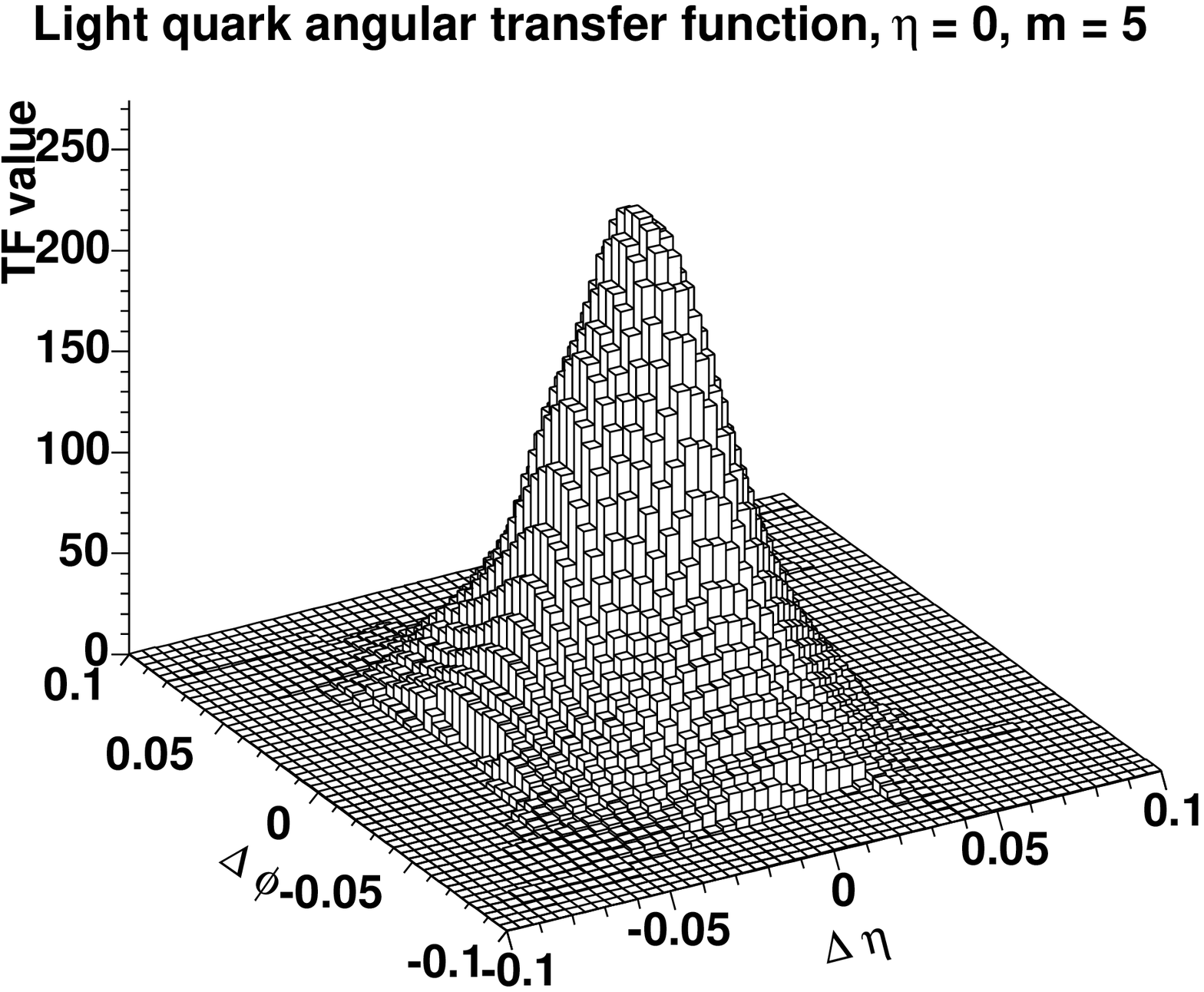}
&	\includegraphics[width=6.8cm, clip=]{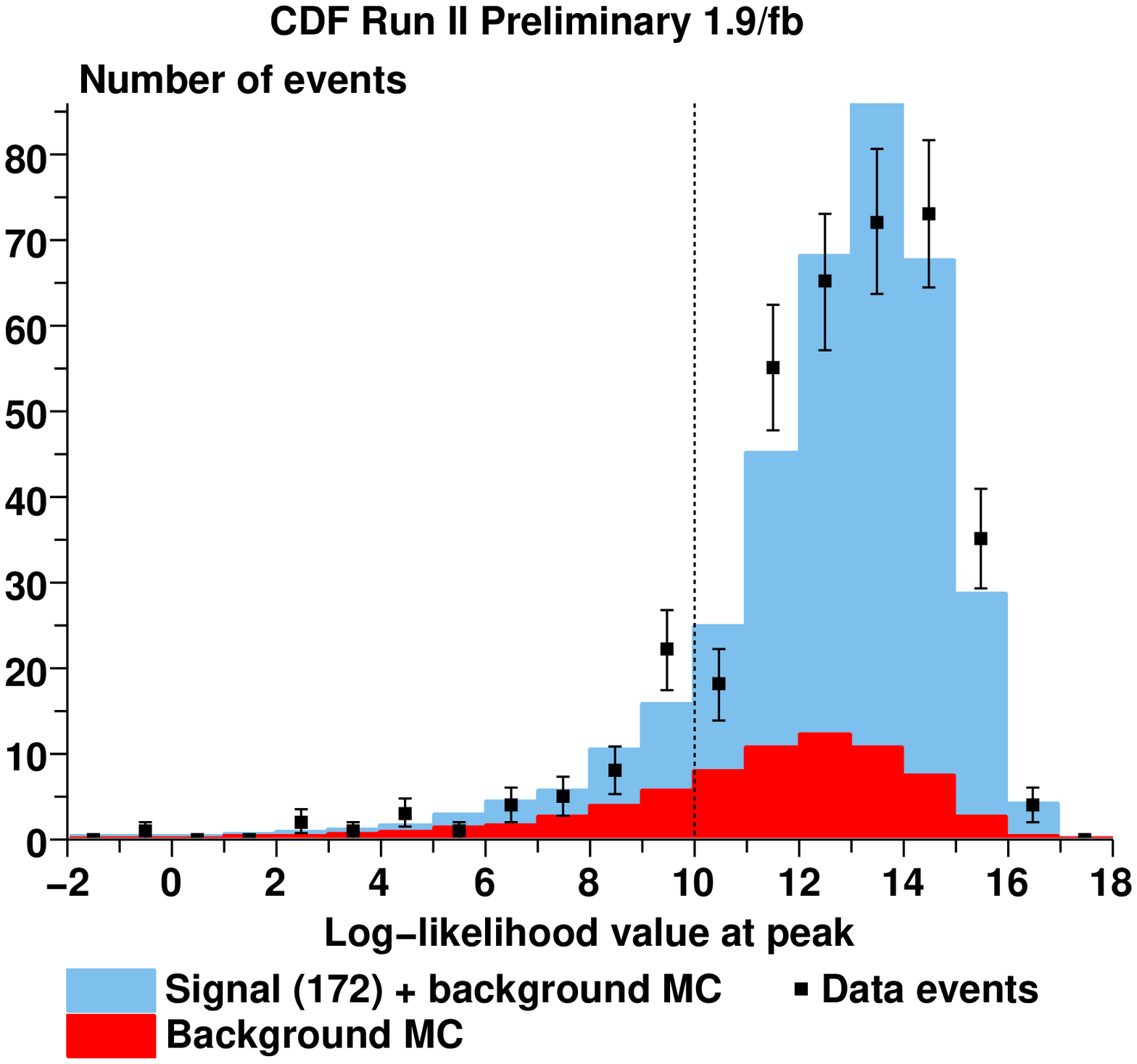}
	\end{tabular}
\caption{(left) The angular transfer functions expressing the distribution of the difference in azimuth angle, $\Delta \phi$, and pseudorapidity, $\Delta \eta$, between the parton and its corresponding jet. (right) A comparison betrween data and Monte Carlo of the distribution of the log-likelihood value corresponding to the maximum of each event likelihood curve.}\label{matrix}
\end{center}
\end{figure}
Matrix element methods define in general a likelihood for each event based on the differential cross section per unit phase space volume of the final state partons, as function of the top quark mass estimator, $M_{top}$. As such they attempt to extract the mamximum amount of kinematic information out of each event, at the expense of a computationally challenging inegration procedure over all unknown parton level quantities. A generic signal event likelihood expression can have the following form:
\begin{equation}
L_{sig}({\bf x};m_t)=\frac{1}{\sigma(m_t)}\int d^n\sigma({\bf y},m_t)dq_1dq_2f(q_1)f(q_2)W({\bf x},{\bf y}),
\end{equation}
where $d^n\sigma({\bf y},m_t)$ stands for the differential cross section of the signal, containing its squared leading order matrix element, $f(q_{1,2})$, the parton density function of the incoming partons, and the transfer function $W({\bf x},{\bf y})$ the probability that a parton level set of variables ${\bf y}$ will be measured as a set of reconstructed event variables, ${\bf x}$. The expresion is ingerated over the unknown quantities $q_1$, $q_2$ and the set ${\bf y}$ and normalised to the total cross section, $\sigma(m_t)$ which is in itself top mass dependent. This technique allows easily the introduction of a free jet energy scale parameterinside the transfer function, $W$. 

The extracted top quark mass estimator can be significantly biased and is generally calibrated using large Monte Carlo samples with different generated top 
quark masses. Also the linearity of the JES measurement is usually verified. Additional background terms are usually added in the event likelihood according to 
varying degrees of sophistication. Some analyses also apply a final selection cut, removing events with the lowest likelihood values.

The most sensitive top mass analysis in CDF~\cite{berkely} applies the matrix element method to a sample of selected $\rm{t\bar{t}}$ candidate events from the 
2fb$^{-1}$ data set. In addition to the kinematical selections and the requirement to have at least one jet tagged as a b-quark jet, this analysis constructs an event-by-event background probability based on a discriminating variable obtained with a neural network. The signal likelihood contributions weighted by this background probability are averaged over all events and subtracted from the final sample log-likelihood.
The signal likelihood itself takes into account all possible jet combinations by which two top quarks can be formed, but weighs them with their respective 
compatibility using b-tag information. In addition to the standard transfer functions that relate the absolute size of the transverse momentum of all partons to their correspinding jets, this analysis also constructs a separate transfer function for the jet engles. An in-situ JES measuerment is possible by including JES information in the transfer functions.
Finally, events with a maximum log-likelihood value below 10 are rejected ince they are highly likely to originate from badly reconstructed signal events.
The angular transfer function and a data-MC comparison of the final peak log-likelihood values of all events are shown in Fig.~\ref{matrix}. 
The measured top quark mass, using this technique equals $M_{top}=171.4 \pm 1.5 (stat. + JES) \pm 1.0 (syst.) GeV/c^2$.

\begin{figure}[t]
\begin{center}
	\begin{tabular}{cc}
	\includegraphics[width=6.8cm, clip=]{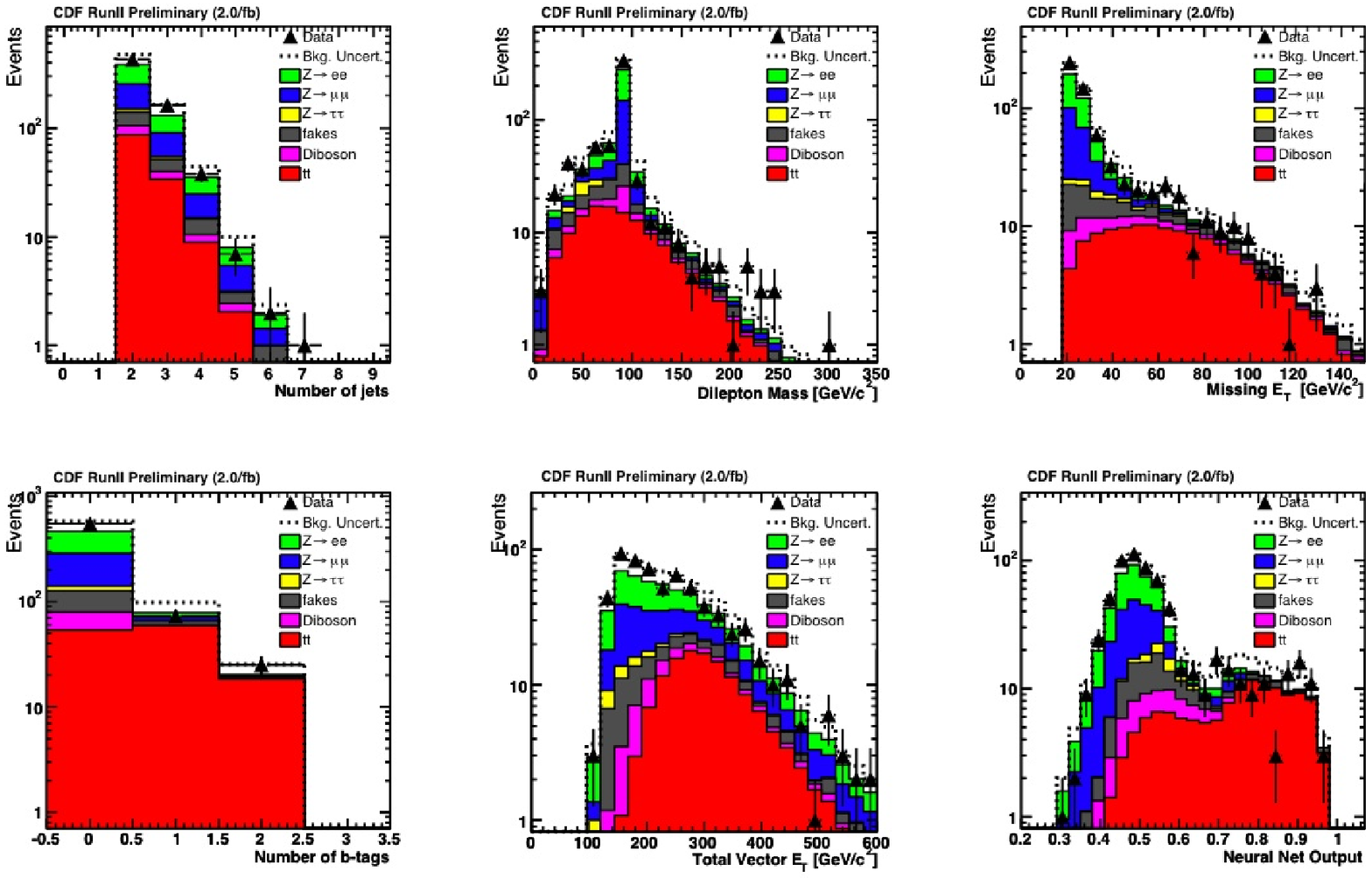}
&	\includegraphics[width=6.8cm, clip=]{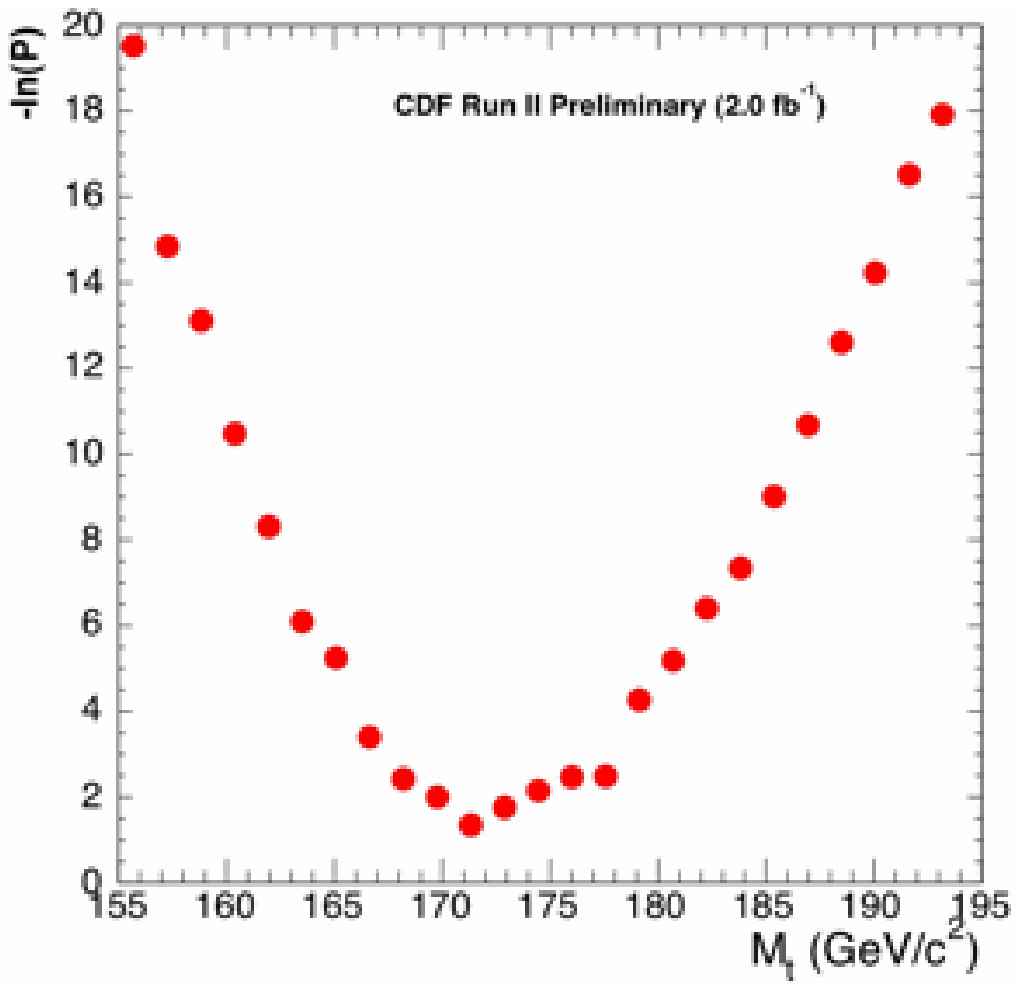}
	\end{tabular}
\caption{(left) The output of the optimized neural network for the dilepton top mass analysis. (right) The negative log-likelihood curve of the top quark mass estimator for the 2fb$^{-1}$ dilepton dataset. }\label{dilme}
\end{center}
\end{figure}

A matrix element method is also applied to the dilepton final state~\cite{daniel}. In this case the event selection is based on an evolutionary neural network that is optimized to yield the smallest expected statistical uncertainty on teh measured top quark mass. The topology and weights within a random collection of neural networks using a set of pre-defined kinematical event variables is allowed to mutate in each optimisation cycle, after which bad performers are rejected and good performers are combined in even more powerful variants. A full ecription of this procedure is given in ~\cite{whitesons}. Another strong feature of this analysis is the use of dedicated background matrix elements to construct a separete background likelihood. This analysis has no strong handle on the JES which is the most dominating systematic uncertainty in this channel.
The mmeasured top quark mass, corresponding to the minimum of the log-likelihood curve shown in Fig.~\ref{dilme} corresponds to a value of  
$M_{top}=171.2 \pm 2.7 (stat.) \pm 2.9 (syst.) GeV/c^2$.

\begin{figure}[t]
\begin{center}
	\begin{tabular}{cc}
	\includegraphics[width=6.8cm, clip=]{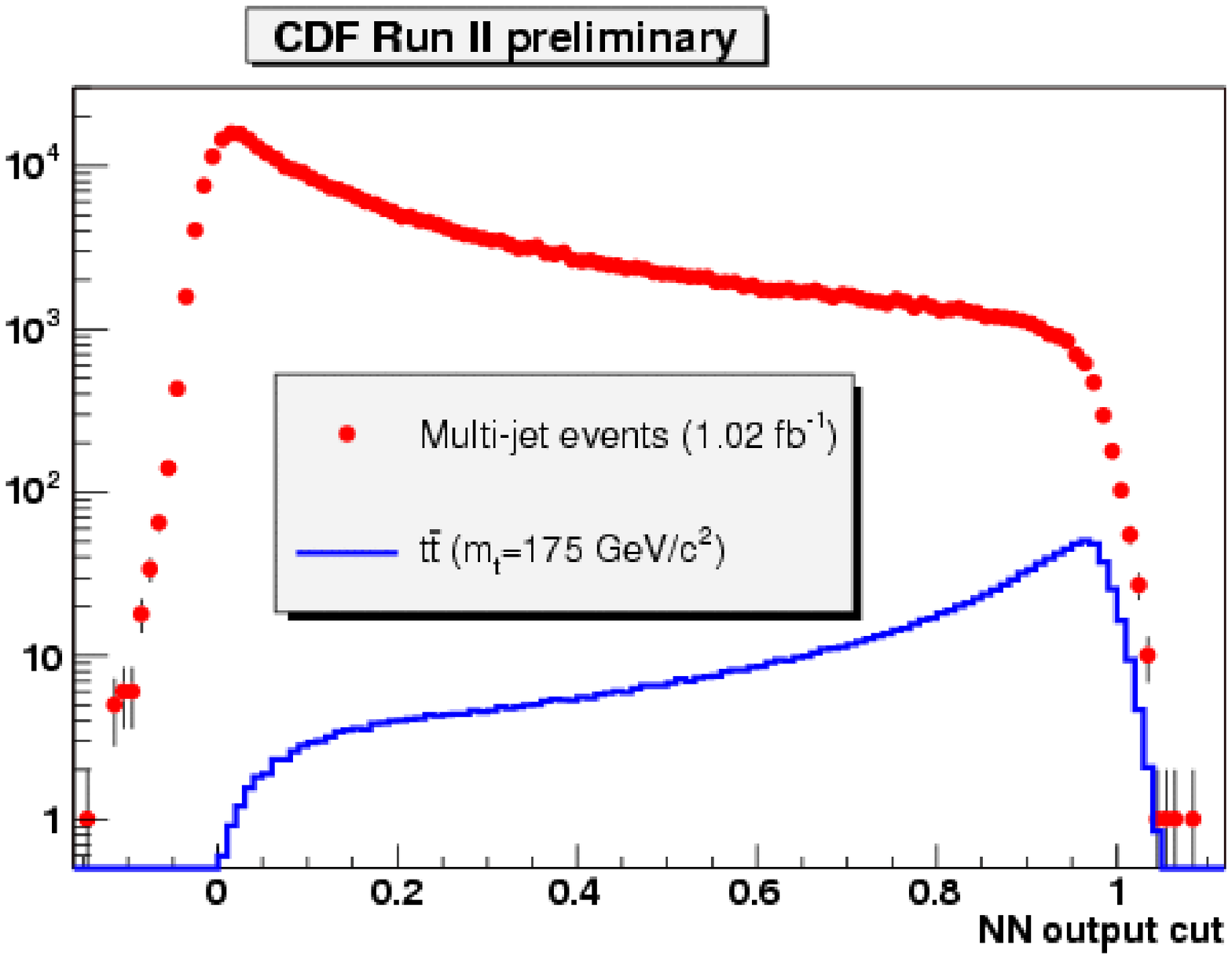}
&	\includegraphics[width=6.8cm, clip=]{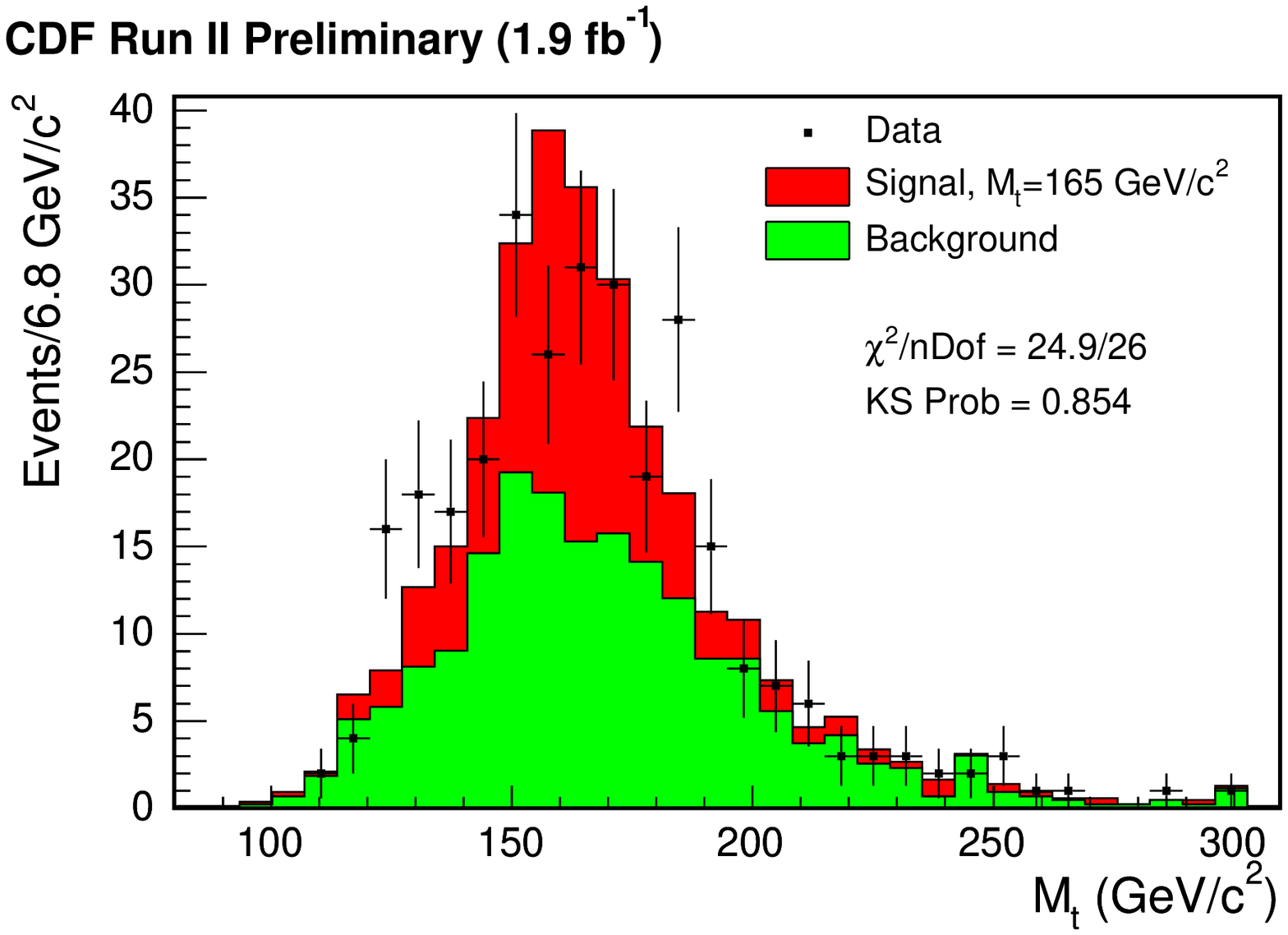}
	\end{tabular}
\caption{(left) The output of the all-hadronic neural network selection, indicating the signal and multijet-QCD constributions. (right) The distrubution of the top quark mass corresponding to the maximum of the signal likelihood in each event.}\label{allhad}
\end{center}
\end{figure}

\section{Ideogram analysis}
The top quark mass is also measured in the all-hadronic channel using a template method~\cite{luca} and a technique known as the ideogram method~\cite{us}.
Both analyses have a common event selection based an a neural network~\cite{allhadsel} of which the output variable is shown in Fig.~\ref{allhad}. In order to increase the signal-to-background ratio to an acceptable value of 2:3, the ideogram analysis requires exactly 6 jets in the final state of which at least two are tagged as b-jets. The event likelihood of the ideogram analysis is based on the decay part of the matrix element. The signal likelihood is in its simplest form
\begin{eqnarray}
L_{sig}(m_t)= &\sum_{i=1}^{90}w_i\int dm'_t dm'_WG(m'_t,m'_W;m_t^i,m_W^i,\sigma_t^i,\sigma_W^i,\rho_{t,W}^i)\\
& \cdot BW(m'_t;m_t,\Gamma_t)BW(m'_W;m_W,\Gamma_W), \notag
\end{eqnarray}
where all 90 possible jet permutations are in principle taken into account, but weighted by a factor $w_i$ that expresses the compatibility of the six final state jets with a decaying $\rm{t\bar{t}}$ pair, using b-tag information and a $\chi^2$ fit that is very similar than the one described in section~\ref{sect:templ}. The integral consists of the convolution of two Breit-Wigner shapes for the top quark and the W boson, with a two-dimensional Gaussian resolution function, $G$, based on the reconstructed top quark and W boson masses in each event, with a proper correlation term, $\rho_{t,W}$, and respective variances, $\sigma_{t,W}$. In case of an in-situ JES measurement, both the weight factor, $w_i$, and the Gaussian resolution function will depend on $\Delta_{JES}$. The ideogram analyis also measures the sample purity in addition to the top quark mass and the JES, yielding results that are compatible with direct cross section measurements. The measured top mass using this technique equals $M_{top}=165.2 \pm 4.4 (stat. + JES) \pm 1.9 (syst.) GeV/c^2$.

\section{Systematic uncertainties}
In most top mass analyses, systematic uncertainties have become the dominant factor in the total uncertainty.
Since more than one year, CDF is revising al of its systematic uncertanty estimates on the emasured top quark mass, in order to have absolute confidence in the small numbers that are quoted, to remove possible double counting between several sources and to carefully study new physics effecst that can be new sources of systematic uncertainty. The current list of systematic sources consists of the uncertainties in the JES (for analysis that are not performing an in-situ measurement of this quantity, initial and final state radiation, uncertainties in the JES for b-quark jets, residual effects of the JES due to the non linear propagation of all correction uncertainties, uncertainties in the parton density functions, mismodeling and generator effects, the effect of multiple interactions, imperfect knowledge of background shapes and fractions and the energy scale of reconstructed leptons. 

Sources that are currently not included, but being investigated, are differences between NLO and LO generators and the use of NLO PDF's and the effect known as color reconnection: a rearrangement of the color flow between final state quarks and between the final state of the hard interaction and the proton remnants. Very promising models that deal with this effect became recently available~\cite{skands} and first attempts are made to generate fully simulated and tuned samples within CDF.

\section{Conclusion}
The latest Tevatron combination~\cite{tevcomb} yields a world average of the top quark mass of $M_{top}=172.6 \pm 0.8 (stat.) \pm 1.1 (syst.) GeV/c^2$.
This implies that we have entered the era where we are truly performing a precison measurement of the top quark mass at the Tevatron. The integrated luminosities exceeding 2 fb$^{-1}$ no longer make the precision statistically limited and most of the systematical uncertainties need to be carefully revised in the near future. Due to this, the LHC experiments will have a hard time to beat the Tevatron precsion in these measurements, since they will face the same challenges of understanding all systematic uncertainties to the sub-percent level. It is also interesting to notice that in parallel, alternative methods that have less sensitivity to the top quark mass but suffer less from systematic uncertainties are being developed and have been proven to work using Tevatron data. The large event yields expected at the LHC wil make these methods very competitive with the mainstream techniques.
Let us not forget that besides serving as a consistency check for the Standard Model, in case of forthcoming evidence, or absence of evidence, for a light neutral Higgs boson, the top quark mass will serve as a calibration tool for many measurements using jets at the LHC.

\end{document}